\documentclass[twocolumn,superscriptaddress,secnumarabic,aps,pra,nobibnotes,groupedaddress,numerical,floatfix,eqsecnum]{revtex4-1}

\usepackage[colorlinks]{hyperref}
\usepackage{graphicx}
\usepackage{float}
\usepackage[squaren]{SIunits}
\usepackage{verbatim}
\usepackage{mathrsfs}
\usepackage{comment}
\usepackage{braket}
\usepackage{bbold}
\usepackage{amsmath}
\usepackage{pdfpages}
\usepackage[normalem]{ulem}
\usepackage{xcolor}
\usepackage{times}
\usepackage{amsmath}

\begin{document}

\title{Frustrated polaritons}

\date{\today}

\author{Sebastian Schmidt}
\affiliation{Institute for Theoretical Physics, ETH Zurich, 8093 Zurich, Switzerland}

\begin{abstract}
Artificially engineered light-matter systems constitute a novel, versatile architecture for the quantum simulation of driven, dissipative phase transitions and non-equilibrium quantum many-body systems. Here, we review recent experimental as well as theoretical works on the simulation of {\it geometrical frustration in interacting photonic systems out of equilibrium}. In particular, we discuss two recent discoveries at the interface of quantum optics and condensed matter physics: (i) the experimental achievement of bosonic condensation into a flat energy band and (ii) the theoretical prediction of crystalline phases of light in a frustrated qubit-cavity array. We show that this new line of research leads to novel and unique tools for the experimental investigation of frustrated systems and holds the potential to create new phases of light and matter with interesting spatial structure. 
\end{abstract}

\pacs{}

\maketitle

\section{Introduction}
 \label{sec:Introduction}

Einsteins postulate of the quantized nature of electromagnetic radiation \cite{Einstein1905} stands at the beginning of quantum physics and paved the way for the discovery of the photon \cite{Compton1923}. Photonic technologies impact our everyday life from the usage of lasers, light-emitting diodes and solar cells to optical fibre communications and diagnostic tools in medicine and the life sciences. In most of these applications photons interact weakly with matter at low light intensities such that a semiclassical description of the electromagnetic fields is still justified. Today, we are in the middle of a new quantum revolution in the field of photonic sciences \cite{Schleich2011}, where it becomes possible to create, store and manipulate single photons on demand in various cavity/circuit QED architectures \cite{Walther2006,Wallraff2004}. When single or few photons interact strongly with well controlled electronic degrees of freedom, the regime of strong light-matter coupling can be accessed with an unprecedented range of novel applications in metrology, sensing and quantum simulation.\\

Quantum simulation, in particular, denotes a concept for solving computationally complex problems in quantum physics and quantum chemistry by utilising artificial quantum machines rather than classical computers \cite{Feynman1982}.
Several platforms for quantum simulation have been suggested, e.g., atomic quantum gases \cite{bloch2012}, trapped ions \cite{Blatt2012}, and photonic systems \cite{Guzik2012}. Photonic quantum simulators based on interacting light-matter systems \cite{Houck2012,Carusotto2013,Schmidt2013*2} are considered as ideal platforms to study non-equilibrium dynamics of open many-body systems such as universality classes of driven dissipative phase transitions and strongly correlated states of photons. It is thus a fascinating challenge from a technological as well as intellectual perspective to think about interesting questions that can be addressed with small scale photonic quantum simulators and how to engineer an efficient and scalable design for such quantum devices.\\

A particular challenging problem of many-body physics concerns the simulation of frustrated systems \cite{Ramirez1994,Moessner2006}. Frustration refers to the impossibility of simultaneously satisfying all possible constraints of a Hamiltonian, e.g., given by the nature of the interactions or the underlying spatial geometry.
As a result, a largely frustrated system is associated with a macroscopic set of degenerate states, i.e., flat energy bands. The many ways in which such a degeneracy can be lifted by residual and often competing interactions gives rise to rich and fascinating physics, e.g., in spin liquids \cite{Balents2010} and spin ice \cite{Ramirez1999,Bramwell2001}. On the other hand, frustrated systems are notoriously difficult to simulate on a classical computer and thus also provide an important testbed for quantum annealers, i.e., adiabatic quantum computers, which are already commercially available \cite{Kadowaki1998,Dasarnab2008,Ronnow2014,Denchev2015}.\\

Artificial lattices exhibiting flat bands were recently implemented using cold atomic gases \cite{Jo2012,Aidelsburger2015}, trapped ions \cite{Kim2010}, Josephson junctions \cite{Sigrist1995,Doucot2002,Feigelman2004}, plasmons \cite{Nakata2012} as well as laser \cite{Nixon2013} and waveguide arrays \cite{GuzmanSilva2014,Vicencio2015,Mukherjee2015}. In this article, we will review recent experimental and theoretical works on geometric frustration in interacting light-matter systems such as exciton-polaritons in nanophotonic band-gap structures \cite{Masumoto2012,Jacqmin2014,Baboux2016} and superconducting qubits embedded in microwave circuitry \cite{Petrescu2012,Biondi2015,Casteels2015}. The open nature of these hybrid systems offers new tools for the experimental investigation of geometric frustration. For example, elementary excitation spectra and spatial/temporal correlation functions can be easily observed in a non-invasive way by analysing the far-field emission of photons using interferometric techniques. On the other hand, the bottom-up approach to the design of artificially engineered light-matter systems offers the possibility to realize a wide range of photonic interactions from weakly correlated systems in the mean-field regime to the regime of strong correlations \cite{Carusotto2013,Schmidt2013*2}.\\

Below, we briefly outline the structure of this topical review: In chapter 2, we give a short but rather general introduction into the field of geometric frustration, from spins and magnets to bosonic systems. In chapter 3, we discuss a particular example of a frustrated, tight-binding lattice exhibiting a flat band, i.e., the Lieb lattice. In chapter 4, we review the recent experimental achievement of bosonic condensation into the flat band of a quasi-1D Lieb lattice, where frustration manifests
as a fragmentation of the condensate due to its strong sensitivity with respect to disorder \cite{Baboux2016}. In chapter 5, we discuss a recent proposal for the realization of an  incompressible state of photons in a frustrated qubit-cavity array \cite{Biondi2015}. Here frustration manifests in a boost of photonic interactions
leading to a novel state of light at the onset of crystallization. A brief summary and outlook is given in chapter 6.

\section{Geometric frustration}

\begin{figure}[t]
\centering
\includegraphics[width=0.4\textwidth,clip]{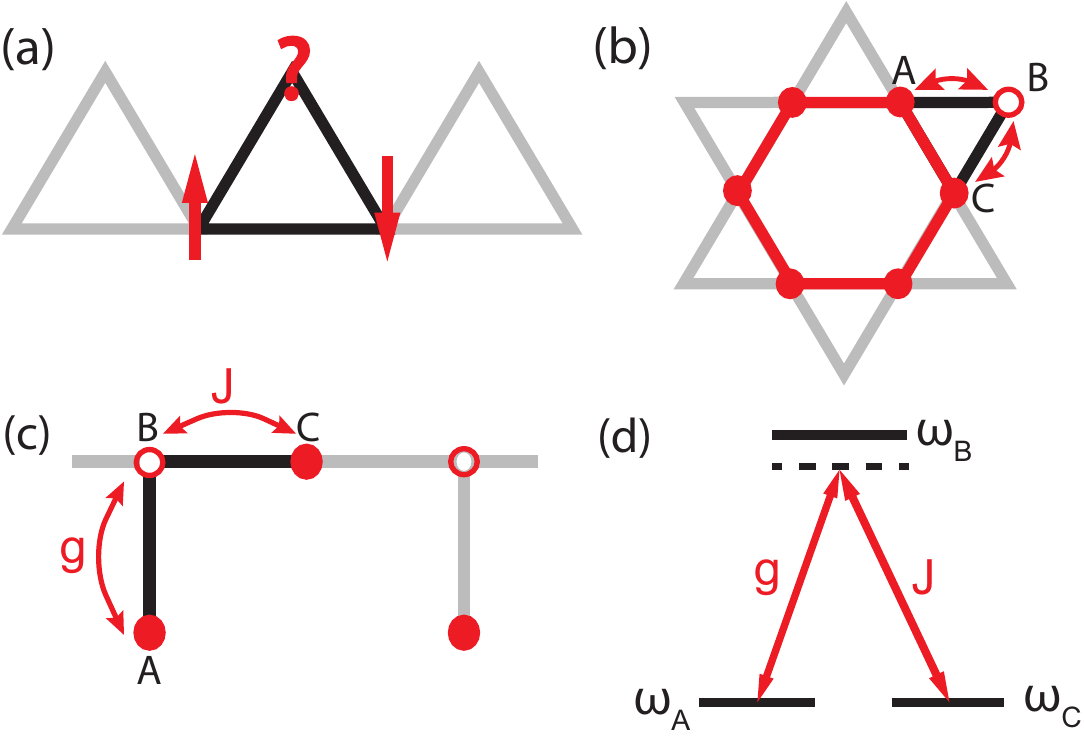}
\caption{(a) Geometric frustration for spins on a triangular lattice due to anti-ferromagnetic coupling. (b) Geometric frustration for bosons hopping on a Kagome lattice due to destructive interference between two hopping paths resulting in localized, hexagonal plaquette states. (c) Same as (b) but for the Lieb lattice. Here, the two paths with hopping strength $g$ and $J$ destructively interfere at site B, which remains dark. This results in a flat band consisting of localized plaquette states, where a single exitation spreads over a C-site and two neighbouring A-sites, see Eq.~(\ref{lambdastate}). (d) Analogous lambda scheme reminiscent of electromagnetically induced transparancy (EIT) (see also text). } 
\label{fig:frustration}
\end{figure}

Geometrical frustration can be of magnetic as well as kinetic origin. Both possibilities are shown in Fig.~\ref{fig:frustration}(a-c). 
Magnetic frustration originates in anti-ferromagnetic spin exchange interactions. The simplest example of magnetic frustration is shown in Fig.~\ref{fig:frustration}(a) with three spins situated at the corners of a triangle \cite{Kim2010}. In this case, two out of three spins antialign, but it is not possible to antialign the third spin with respect to the other two. Consequently, not all antiferromagnetic interactions can be satisfied simultaneously as they are incompatible with the triangular lattice symmetry. The ground-state of three spins on a triangle will then be sixfold degenerate corresponding to all possible combinations with two antiferromagnetic and one ferromagnetic bond. With larger lattice sizes, the degeneracy also quickly increases. In real materials this degeneracy is at least partially lifted due to local magnetic fields or unequal coupling strength leading to distorted lattices. A hallmark of anisotropic frustrated lattices is the stepwise lifting of the degeneracy as a function of an externally applied magnetic field such that the magnetization describes a characteristic staircase function \cite{Honecker2004,Ueda2005,derzhko2015}. The large degeneracy of the ground state also leads to a finite entropy at zero temperature in analogy with water ice (where a configurational disorder inherent to the protons in the water molecule leads to a residual low temperature entropy \cite{Pauling1935}). Two and three-dimensional spin lattices, which exhibit geometrical frustration are therefore referred to as spin ice (for a review see \cite{Bramwell2001,Moessner2006}). The first natural spin ice materials were discovered in 1997 and belong to the group of rare earth pyrochlores \cite{Harris1997}.\\

In bosonic systems, macroscopically degenerate ground states arise due to frustrated hopping in certain tight-binding lattices \cite{Bergman2008,Huber2010}. Some generic criteria for the existence of flat bands can be found in Refs. \cite{Lieb1989, Mielke1991,Tasaki1992,Bergman2008,Chalker2010}.
For example, the unit cell of a Kagome lattice shown in Fig.~\ref{fig:frustration}(b) offers two possible path for a particle to hop on one of the corners of the Kagome star, i.e., it can hop from site A or C to site B. These two path can destructively interfere leading to a completely localized eigenstate that lives only on the inner hexagon. Such localized plaquette states then form a special single-particle band for the whole lattice, which is completely flat, i.e., dispersionless, in the entire Brillouin zone (the degeneracy of the single-particle flat band is given by the number of hexagons in the lattice). In principle, such bosonic flat band systems can be engineered for ultra-cold atoms using optical lattices. However, experimental studies of frustrated ground-states (where the flat band is the lowest band) are difficult as this requires complex-valued hopping constants \cite{Aidelsburger2015}. Alternatively, one might use special pump schemes to coherently transfer atomic condensates into an excited flat band \cite{Taie2015}.\\ 

In both cases discussed above, frustration leads to macroscopic degeneracies, i.e., a divergence in the density of states. The latter prevents straightforward ordering and leads to an extreme sensitivity of the system with respect to small perturbations. For example, disorder can lead to the phenomenon of inverse Anderson localization \cite{Goda2006} and unconventional critical exponents \cite{Chalker2010,Bodyfelt2014}. On the other hand, repulsive interactions lead to strongly correlated phases of matter such as resonance-valence bonds \cite{Anderson1987}, charge-density waves \cite{Huber2010} and Wigner crystallization \cite{Guanyu2014}. In particular, competing interactions typically result in rich phase diagrams and exotic physics \cite{Balents2010}. 
An important motivation for the study of frustrated systems thus originates in material science, e.g., to gain a better microscopic understanding of high-Tc superconductors \cite{Emery1993,Sigrist1995}. The degeneracy of frustrated systems is also closely analogous to the quantum Hall effect, where a 2D electron gas subject to an external magnetic field exhibits macroscopically degenerate Landau levels. Also here, the many ways in which this degeneracy can be lifted leads to highly non-trivial ground states \cite{Laughlin1983,Neupert2011,Tang2011}. \\

Due to the plethora of rich and interesting physics on one hand and the difficulty of simulating large, degenerate systems on a classical computer, frustrated systems are an ideal candidate for quantum simulations \cite{Buluta2009,Kim2010}.

\section{Lieb lattice}

In the following, we will consider a particularly simple example of a lattice featuring a flat band, i.e., the so-called Lieb lattice whose unit cell is shown in Fig.~\ref{fig:frustration}(c). A quasi-1D cut through the Lieb lattice (also referred to as stub lattice) is described by the tight-binding Hamiltonian
\begin{eqnarray}
\label{lieb}
H_{\rm Lieb}=\sum_{j=1}^N h_{j} + J\sum_{j=1}^{N-1}\left[\left(b_j + b_{j+1}\right)c_j^\dagger + {\rm h.c.}\right]
\end{eqnarray}
with the on-site Hamiltonian
\begin{equation}
h_j=\omega_{\rm \scriptscriptstyle A} a_j^\dagger a_j +  \omega_{\rm \scriptscriptstyle B} b_j^\dagger b_j +  \omega_{\rm \scriptscriptstyle C} c_j^\dagger c_j + g\left(a_j^\dagger b_j + {\rm h.c}\right)\,,
\end{equation}
where the bosonic operators $a_j, b_j, c_j$ annihilate a particle at site $A,B,C$ in unit cell $j=1,...,N$. The second term in (\ref{lieb}) describes particle
hopping between nearest neighbour sites at a rate $J$. 
The Hamiltonian (\ref{lieb}) is conveniently diagonalized using the Fourier transform $\boldsymbol{\psi}_{j} = (1/\sqrt{N})\sum_k e^{ikj} \boldsymbol{\psi}_{k}$ with $\boldsymbol{\psi}_{j} = [\,a_{j},\,b_{j},\,c_{j}\,]^T$. By imposing periodic boundary conditions, we obtain the $k$-space representation of the lattice Hamiltonian, i.e.,
\begin{equation}
\label{fourier}
H_{\rm Lieb} = \sum_{k} \boldsymbol{\psi}_{k}^\dagger \left[ \begin{array}{ccc} \omega_{\rm \scriptscriptstyle A} & g & 0 \\  
g & \omega_{\rm \scriptscriptstyle B} & J(1+e^{-ik}) \\
0 & J(1+e^{ik}) & \omega_{\rm \scriptscriptstyle C} \end{array} \right] \boldsymbol{\psi}_{k}
\end{equation}
with $k = 2\pi n/N$, $n = -N/2,\dots, N/2-1$.
The diagonalization of the single-particle Hamiltonian in (\ref{fourier}) yields in general three dispersive bands. 
If $\omega_{\rm \scriptscriptstyle A}=\omega_{\rm \scriptscriptstyle C}$ the middle band turns flat with energy 
\begin{eqnarray}
\label{flatenergy}
\varepsilon_{\rm \scriptscriptstyle FB} = \omega_{\rm \scriptscriptstyle A}=\omega_{\rm \scriptscriptstyle C}\,,
\end{eqnarray}
while the other two remain dispersive with energies 
\begin{eqnarray}
\label{dispband}
\varepsilon_{k}^\pm  = \omega_{\rm \scriptscriptstyle C} +\delta_{\rm \scriptscriptstyle BC}/2 \pm \sqrt{2J^2(1 + \cos k)+ g^2 + \delta_{\rm \scriptscriptstyle BC}^2/4}\nonumber\\
\end{eqnarray}
and $\delta_{\rm \scriptscriptstyle BC}=\omega_{\rm \scriptscriptstyle B} - \omega_{\rm \scriptscriptstyle C}$.
Note, that the flat band energy in (\ref{flatenergy}) does not depend on $g$ or $J$ and is separated by a gap $\sqrt{g^2 + \delta_{\rm \scriptscriptstyle BC}^2/4} \pm \delta_{\rm \scriptscriptstyle BC}/2$ from the dispersive bands (\ref{dispband}). The flat band eigenstates can be written as
\begin{equation}
\ket{\Lambda_{j}} = \Lambda_{j}^\dagger \ket{\text{vac}}
\label{lambdastate}
\end{equation}
with the flat band operator 
\begin{equation}
\Lambda_{j}^\dagger = \frac{1}{\sqrt{g^2 + 2J^2}}\big[ g\,c_{j}^\dagger -J(a_{j}^\dagger + a_{j+1}^\dagger)\big],
\end{equation}
which describes a localized plaquette defined by one C and two neighbouring A sites (see filled circles in Fig.~\ref{fig:frustration}(c)). The flat band arises due to the destructive interference between a photon hopping from resonator C to B ($\sim J$) and a photon hopping process between site A and B ($\sim g$). As a consequence the B cavities remain completely dark.\\

The matrix in (\ref{fourier}) is formally identical to a driven three-level system in the so-called $\lambda$-configuration \cite{Fleischhauer2005} as depicted in Fig.~\ref{fig:frustration}(d), where the three levels are equivalent to a single photon residing on site $A$, $B$ or $C$. In this picture, the hopping rates $g$ and $J$ are equivalent to two coherent drives exchanging excitations between level $B$ and $A$ as well as $C$ with drive strength $g$ and $J$. It is well known that such a system features a dark state if the energies of the lowest two levels are identical, i.e., $\omega_{\rm \scriptscriptstyle A}=\omega_{\rm \scriptscriptstyle C}$. Under this condition, level $B$ remains empty due to the destructive interference between the two drives in analogy to the plaquette state in (\ref{lambdastate}). In quantum optics, large ensembles of such three-level emitters are used to slow down the propagation of light in a scheme known as electromagnetically-induced transparancy (EIT) \cite{Fleischhauer2005}. In analogy to slow light propagation in an EIT medium, an excitation originally localized at one end of the Lieb chain in Fig.~\ref{fig:frustration}(c) does not disperse and/or propagate to the other end.\\

The Lieb lattice has recently been realized using waveguide arrays, where the single-particle dispersion including the flat band in (\ref{flatenergy}) and its dark states in (\ref{lambdastate}) were measured \cite{GuzmanSilva2014,Vicencio2015,Mukherjee2015}. Bosonic condensation into the flat band of the Lieb lattice was realized in Ref. \cite{Baboux2016} using exciton-polaritons confined to an array of semiconductor micro-pillars. The results of this work are discussed in more detail in the next chapter.

\begin{figure}[t]
\centering
\includegraphics[width=0.35\textwidth,clip]{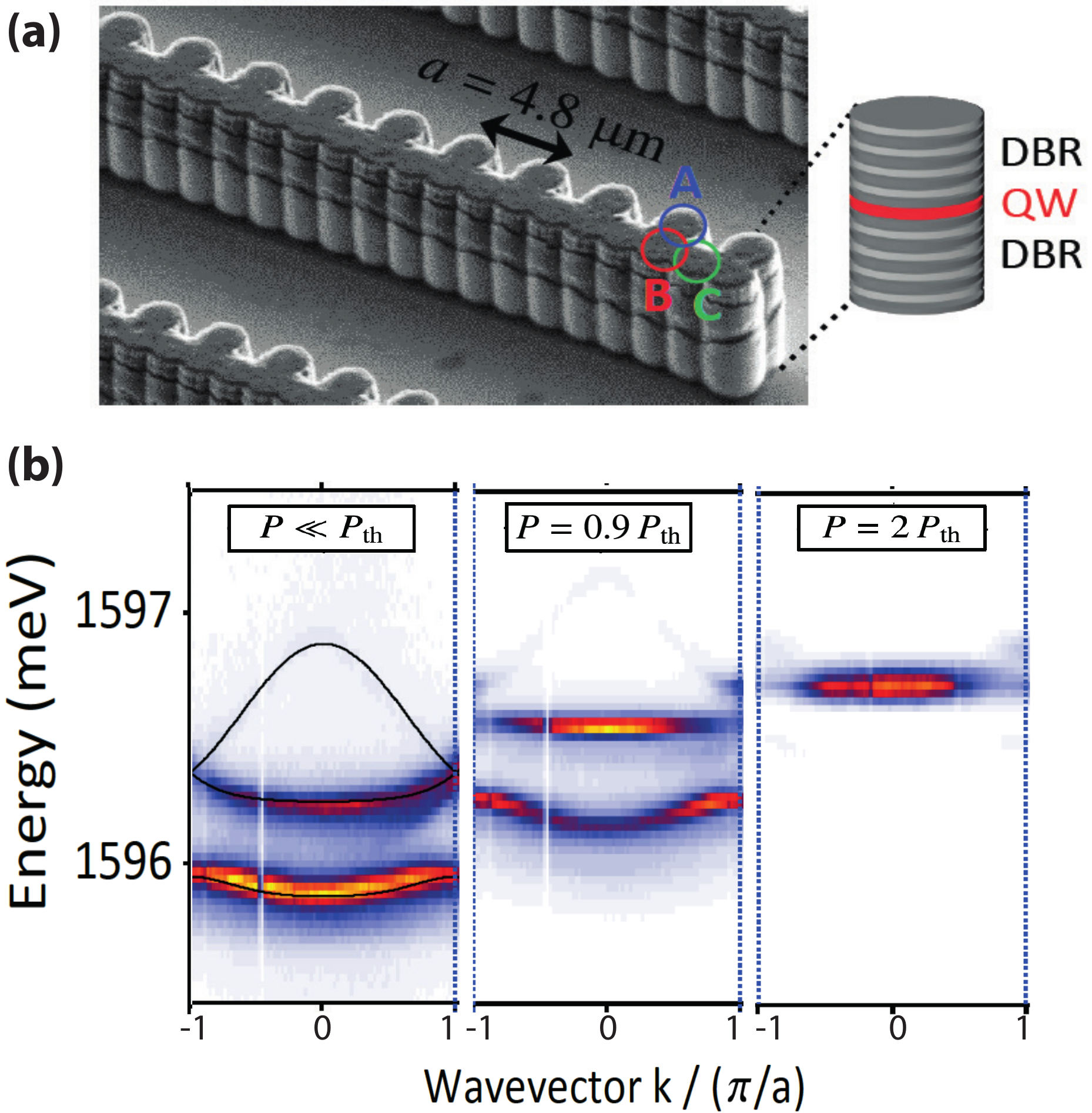}
\caption{(a) Scanning electron micrograph of a 1D Lieb lattice as shown in Fig.~\ref{fig:frustration}(c), made of semiconductor micropillars, each containing a quantum well (QW) sandwiched between two DBR mirrors (cavity). Exciton-polaritons are formed in-plane due to a coherent coupling between cavity photons and quantum well excitons. (b) Energy resolved emission of the array in (a) for different pump strength $P$ for the case of asymmetric pumping (see Fig.~\ref{fig:pump}(d)). Far below treshold ($P\ll P_{\rm th}$), the emission is incoherent and originates from several bands. At treshold ($P=P_{\rm th}$) polaritons condense into the flat band.
As the pump strength increases above treshold ($P=2P_{\rm th}$), emission occurs solely from the flat band. Note, that all bands are blue-shifted by the pump. 
The middle band is almost completely flat just before treshold, because the blue shift of the A pillars in the asymmetric pump configuration depicted in Fig.~\ref{fig:pump}(d) shifts the effective on-site energies of the A-sites into resonance with the C-sites. Figure adapted with permission from Ref. \cite{Baboux2016}. Copyrighted by the American Physical Society.}
\label{fig:pillar}
\end{figure}
\begin{figure*}[t]
\centering
\includegraphics[width=0.7\textwidth,clip]{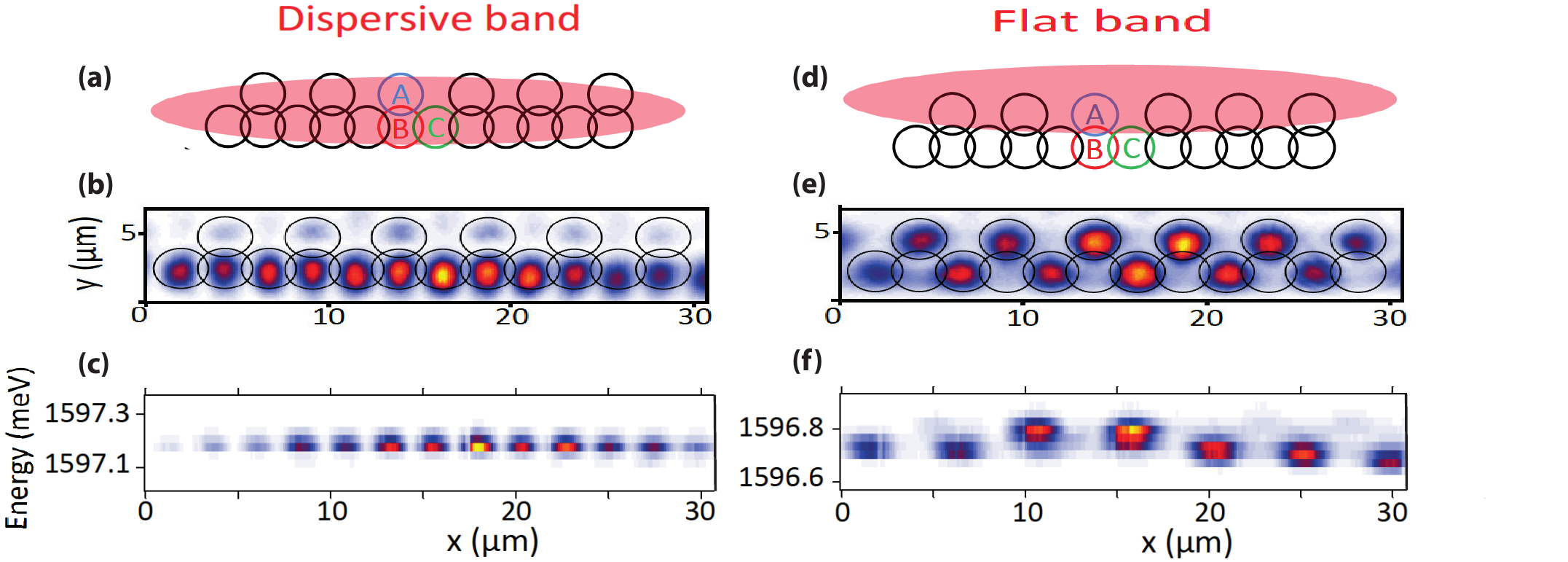}
\caption{(a) Symmetric pump scheme for condensation into a dispersive band of the Lieb lattice in Fig.~\ref{fig:pillar}(a) with the real space emission shown in (b). (c) shows the energy resolved emission of a synchronized condensate emitting at a single frequency (over the whole pump spot). Panels (d)-(f) show the corresponding figures for asymmetric pumping into a flat band. The real space emission in (e) reflects the spatial structure of the plaquette states in (\ref{lambdastate}) with dark B-pillars. The energy resolved emission in (f) demonstrates the fragmentation of the condensate, where each fragment has the size of a plaquette and emits at a different frequency. Figure adapted with permission from Ref. \cite{Baboux2016}. Copyrighted by the American Physical Society.} 
\label{fig:pump}
\end{figure*}

\section{Condensation of light}

In this section, I will give a short introduction on polariton condensates in general, before I discuss the realization of a frustrated condensate in a flat energy band
as reported in \cite{Baboux2016}.\\
Polariton condensates are quantum fluids of light made of exciton-polaritons \cite{Kasprzak2006}, i.e., quasiparticles which form 
due to a coherent coupling between photons in microcavities and excitons confined in low-dimensional semiconductor structures (see Fig.~\ref{fig:pillar}(a)). The resulting polariton bands can be tuned by the spatial profile of the pump laser, which creates excitons and causes a blue shift of the confining potential (see Fig.~\ref{fig:pillar}(b)). The photons emitted by the cavity can be efficiently detected, which gives direct access to the polariton occupation of the bands as well as their spatial/temporal correlation functions. On the other hand, the excitonic component provides effective nonlinearities for (i) efficient polariton-polariton scattering from excited states into the condensate modes and (ii) polariton-polariton interactions within the condensate. The strength of the polariton nonlinearity thus also depends on the pump power, i.e., the number of excitons.\\

In the case of weak, off-resonant pumping, the emission from the cavity is mostly incoherent. 
As the pump strength increases, stimulated polariton-polariton scattering into a specific mode may overcome losses due to photon
leakage and exciton recombination. In the case of condensation, a single mode becomes macroscopically occupied above a certain critical pump strength. Emission from that mode is characterized by temporal as well as spatial coherence similar to atomic BEC`s. However, the occupation of the excited states outside the condensate typically does not obey a thermal distribution but follows from a non-equilibrium steady state which is determined by a balance of drive and dissipation (for an extensive review on polariton condensates, see Ref.~\cite{Carusotto2013}).\\

Recently, polariton condensates were also realized in various micropillar structures, where the confining potential for polaritons is periodically modulated by dry etching techniques (see Fig.~\ref{fig:pillar}(a)). Size and distance of each pillar can be as small as a few micrometers. The spatial overlap of evanescent photonic modes between two pillars then allows for tunneling of polaritons and the realization of tight-binding lattices such as Kagome or Lieb lattices \cite{Abbarchi2012,Jacqmin2014, Baboux2016}.\\

In a generic tight-binding lattice systems, the condensate above treshhold is described by a simple, non-Hermitian Hamiltonian, which is derived from a generalized Gross-Pitaevski equation \cite{Carusotto2013}. For the case of the Lieb lattice as shown in Fig.~\ref{fig:pillar}(a), the Hamiltonian reads
\begin{eqnarray}
\label{GPHamil}
H=H_{\rm Lieb} + \frac{1}{2}\sum_{i}  \Gamma_i   \hat{n}_i +\, {\rm nonl.}\,,
\end{eqnarray}
where $H_{\rm Lieb}$ is given in (\ref{lieb}) and the sum runs over every site $i$ of the lattice. Here, $\hat{n}_i$ denotes the particle number operator on each site (e.g., $n_i=a_i^\dagger a_i$ for an A site {\rm etc}.) and $\Gamma_i$ is a complex-valued rate which includes the effects of incoherent pumping and dissipation. The real part of these rates determines the blue shift of each pillar, i.e.,
\begin{eqnarray}
\label{realGamma}
\operatorname{Re}(\Gamma_i)= 2 P_i\, \frac{g_r}{\gamma_x}\,,
\end{eqnarray}
which is proportional to the pump power $P_i$. This collective blue shift accounts for the repulsive interaction between polaritons inside the condensate and the polaritons outside of the condensate (often approximated as a pure excitonic reservoir with a marginal photonic component). The two-polariton scattering rate for this process is given by $g_r$ and $\gamma_x$ denotes the exciton recombination rate. The imaginary part 
\begin{eqnarray}
\label{imagGamma}
\operatorname{Im}(\Gamma_i)= P_i\, \frac{R}{\gamma_x}  - \kappa
\end{eqnarray}
determines the effective decay rate of a polariton condensate in a cavity with photon loss rate $\kappa$. The first term on the {\rm r.h.s} of (\ref{imagGamma}) denotes the gain, which is proportional to the pump power $P_i$ as well as the exciton relaxation rate $R$ from the reservoir into the condensate.
Note, that the blue shift in (\ref{realGamma}) as well as the effective decay rate in (\ref{imagGamma}) depend indirectly on the number of excitons in the reservoir through the exciton recombination rate $\gamma_x$. In Eq.~(\ref{realGamma})-(\ref{imagGamma}), we have assumed that the rates are identical for each pillar. A spatial dependence of $\Gamma_i$ arises only through the pump profile $P_i$ for simplicity.\\

The eigenmodes and corresponding eigenvalues of the non-hermitian Hamiltonian in (\ref{GPHamil}) are complex. 
If all eigenvalues have a negative imaginary part, the corresponding eigenmodes decay at long times and no stable condensate exists.
This is the case when the pump is weak, i.e., when photonic losses dominate over the gain in (\ref{imagGamma}). With increasing pump strength, the eigenvalues start to move towards the real axis. A critical pump power is reached as soon as one eigenvalue crosses the real axis. In this case the system starts to condense into the corresponding eigenmode. For a single, independent pillar the critical treshhold is obtained from (\ref{imagGamma}) as
\begin{eqnarray}
P_{\rm th} \sim \frac{\gamma_x \kappa}{R}\,.
\end{eqnarray}
Note, that in (\ref{GPHamil}) we have neglected all nonlinear contributions, because the interaction strength of a single pair of polaritons within the condensate is typically much smaller than the polariton loss rate. However, this does not mean that interactions do not play a role. On the contrary, they are important for the condensation process itself, i.e., for a stimulated scattering of polaritons into the condensate. By strong resonant or off-resonant pumping, the {\it collective} Kerr-type interaction within the condensate can exceed the loss rate and may lead to a weakly correlated but nonlinear superfluid \cite{Amo2009}. Even the regime of strong correlations can be accessed in strongly confined excitonic systems such as self-assembled quantum dots \cite{Reinhard2012}. Further details on the derivation and validity of (\ref{GPHamil})-(\ref{imagGamma}) can be found in Ref.~\cite{Carusotto2013}.\\ 

\begin{figure}[t]
\centering
\includegraphics[width=0.35\textwidth,clip]{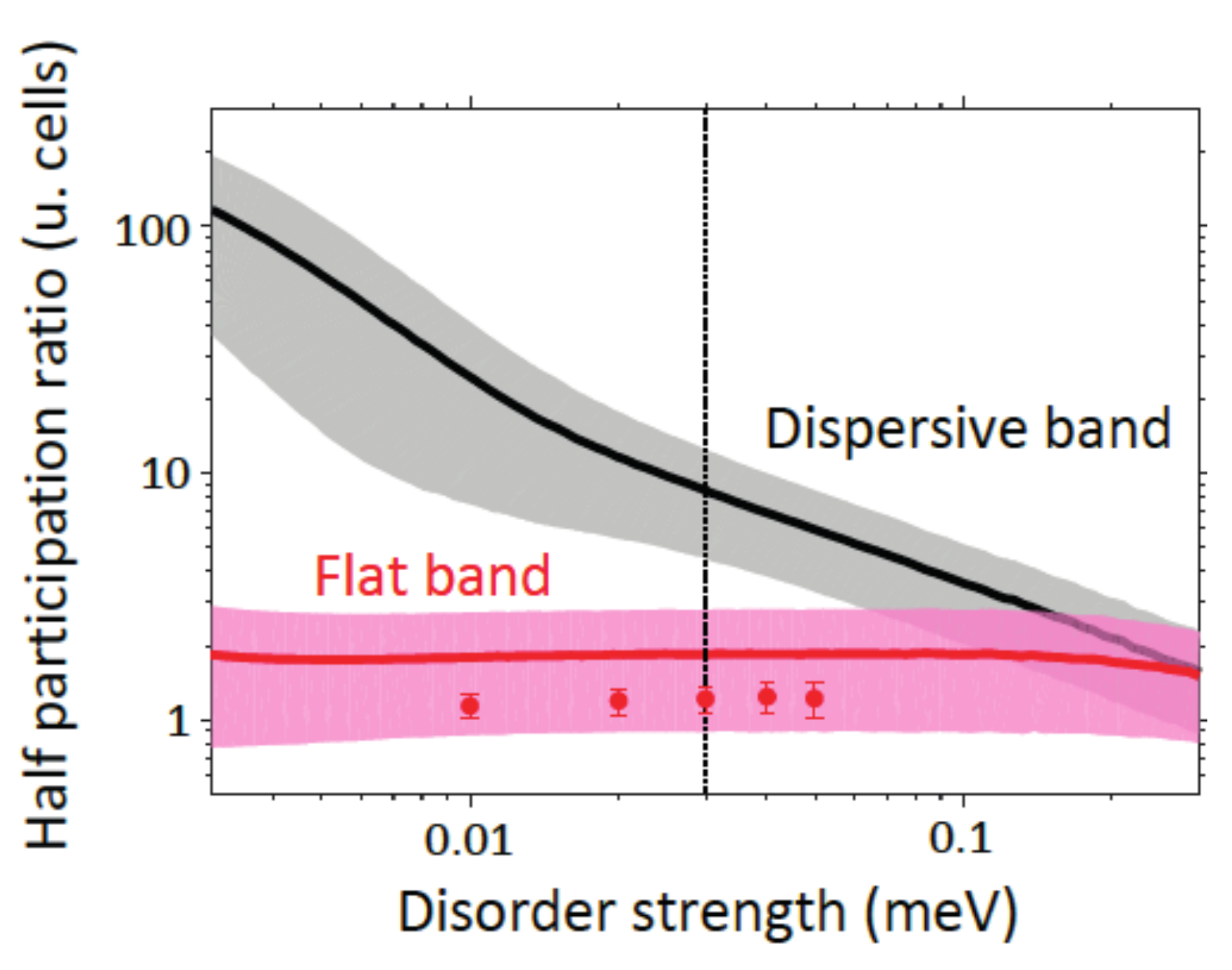}
\caption{Calculated, disorder-averaged half participation ratio ({\rm HPR}) for the quasi-1D Lieb lattice in (\ref{lieb}) with randomly distributed on-site energies drawn from a Gaussian distribution. While the {\rm HPR} in the dispersive band depends strongly on the disorder strength, states of the flat band localize on the order of $1\sim 2$ unit cells even at infinitesimal disorder. The grey shading shows the standard deviation of the ({\rm HPR}), while the circles depict results for the HPR including drive and dissipation. Figure adapted with permission from Ref. \cite{Baboux2016}. Copyrighted by the American Physical Society.} 
\label{fig:hpr}
\end{figure}

How is it possible to trigger condensation into a flat rather than a dispersive band? From (\ref{imagGamma}) it follows that modes, which have support on sites that are strongly pumped will have a large gain and thus a lower condensation treshhold compared to modes which have mostly support on weakly pumped sites. For example, in the case of the Lieb lattice discussed in the previous section the flat band states in (\ref{lambdastate}) have a larger amplitude on the A-sites compared with modes in the dispersive bands. By tailoring the profile $P_i$ of the laser such that it pumps favorably the A-sites of the Lieb lattice, one can thus trigger condensation into the flat band mode. Bosonic condensation into a flat energy band using such a tailored incoherent pump scheme has first been achieved in Ref.~\cite{Baboux2016}. The results of the measurements are shown in Fig.~\ref{fig:pillar}(b) and Fig.~\ref{fig:pump}.
With the symmetric pump profile in Fig.~\ref{fig:pump}(a) the system condenses into the upper dispersive band. However, an asymmetric pump profile as shown in Fig.~\ref{fig:pump}(d) triggers condensation into the flat band. 
The real space image of the emission in Fig.~\ref{fig:pump}(b) shows the condensate of a dispersive band with bright B sites, but little weight on the A sites.
In comparison, Fig.~\ref{fig:pump}(e) demonstrates that the B pillars of the flat band condensate remain dark characteristic of the plaquette states in (\ref{lambdastate}).\\

What is the spatial structure of the flat band condensate? The eigenstates of the flat band can be written in terms of the localized plaquette states in (\ref{lambdastate}). However, since all plaquette states are degenerate, one could just as well form a non-local basis, which consists of a superposition of different plaquettes. Thus, it is a-priori not clear what a flat band condensate should actually look like. The answer to this question is shown in Fig.~\ref{fig:pump}(f) which depicts the energy-resolved emission of each unit cell. It turns out that the flat band condensate in Fig.~\ref{fig:pump}(f) does not extend uniformly over the whole lattice, but rather fragments into smaller condensates with a size on the order of $1\sim 2$ unit cells. Each of these fragments maintains temporal coherence, but emits at a different frequency. Interferometric measurements of the first-order spatial coherence function confirm this picture and show an exponential decay for the visibility of the interference fringes with a characteristic length scale of about $\sim 1.6\pm 0.3$ unit cells. This is in strong contrast to the condensate in the dispersive band shown in Fig.~\ref{fig:pump}(c), where about $\sim 10$ pillars emit at the same energy and are perfectly synchronized \cite{Baboux2016}.\\
Such a markedly different spatial structure of both condensates has its origin in the extreme sensitivity of the flat band with respect to disorder. 
In the case of the micropillars shown in Fig.~\ref{fig:pillar}(a), diagonal disorder originates from small variations of the radius of each pillar and is estimated to be $30\micro eV$, i.e., roughly 15 \% of the hopping strength $g$ and $J$ (and thus much smaller than the gap to the dispersive bands).
Any small amount of (diagonal) disorder in the on-site energies breaks the degeneracy of the flat band and leads to strong localization of the eigenstates. The localization length of a mode can be calculated from the half participation ratio
\begin{eqnarray}
\label{hpr}
{\rm HPR}=\frac{1}{2}\frac{\left(\sum_j |\psi_j|^2\right)^2}{\sum_j |\psi_j|^4}\,,
\end{eqnarray}
where $\psi_j$ denotes the wavefunction amplitude on site $j$ of the lattice. 
Anderson localization theory predicts that all modes of a quasi one-dimensional tight-binding lattice are exponentially localized. Writing the ansatz $\psi_j\sim \exp(-|j|/\zeta)$ and replacing the sum in (\ref{hpr}) by an integral yields ${\rm HPR}=\zeta$ for the case of an infinite lattice. The half participation ration ${\rm HPR}$ thus represents a useful measure for localization and can be calculated in the case of the Lieb chain by diagonalizing
the Hamiltonian in (\ref{lieb}) with on-site energies drawn from a Gaussian distribution with variance $\sigma$. The result is shown in Fig.~\ref{fig:hpr}. For the dispersive band, the ${\rm HPR}$ shows a strong dependence on energy as well as disorder strength. On the contrary, in the case of the flat band, the localization length is almost insensitive with respect to the disorder strength and is roughly given by the size of the smallest plaquette. Consequently, even at infinitesimal small disorder, the flat band modes immediately localize on the size of roughly $1\sim2$ unit cells \cite{Baboux2016}.\\
On the other hand, the decay of the interference fringes as obtained from an interferometric measurement, yields $\zeta\sim 10$ unit cells for the dispersive band and $\zeta\approx 1.6\pm 0.3$ unit cells for the flat band in agreement with the theoretical prediction in Fig.~\ref{fig:hpr}. Note, that Fig.~\ref{fig:hpr} also includes a few results for the {\rm HPR} as obtained from the full Hamiltonian in (\ref{GPHamil}) (i.e., including drive and dissipation), which are close to the calculation using $H_{\rm Lieb}$ alone.\\

To conclude, the fragmentation of the condensate in the flat band is a consequence of the extreme sensitivity with respect to disorder in a frustrated system with small intrinsic nonlinearities \cite{Baboux2016}. In the next section, I will discuss a different regime, i.e., the interplay of frustration and strong interactions in a coherently driven, dissipative light-matter system \cite{Biondi2015}.

\section{Crystallization of light}

The realization of strong light-matter interactions in various cavity/circuit QED systems has triggered an immense interest in realizing condensed phases and strongly correlated states of photons \cite{Houck2012,Carusotto2013,Schmidt2013*2}. 
One of the most fascinating questions in this context is whether one can boost photonic interactions to an extreme regime, where light itself crystallises similarly to a superfluid-Mott insulator (SF-MI) phase transition \cite{Greentree2006,Hartmann2006,Angelakis2007}.\\ 
The Jaynes-Cummings-Hubbard model (JCHM) has been introduced to describe a possible SF-MI transition of polaritons in coupled qubit-cavity arrays under quasi-equilibrium conditions \cite{Greentree2006,Angelakis2007}. 
The competition between repulsive photon interactions (localization) and the photon hopping between cavities (delocalization) leads to an equilibrium quantum phase diagram featuring Mott lobes reminiscent of those of ultracold atoms in optical lattices as described by the Bose-Hubbard model ~\cite{Fisher1989}. This extreme many-body state of light has been the subject of intense theoretical investigations (for an overview on this subject, see the reviews \cite{hartmann2008,Houck2012,Schmidt2013*2} and literature therein).\\
\begin{figure}[t]
\centering
\includegraphics[width=0.45\textwidth,clip]{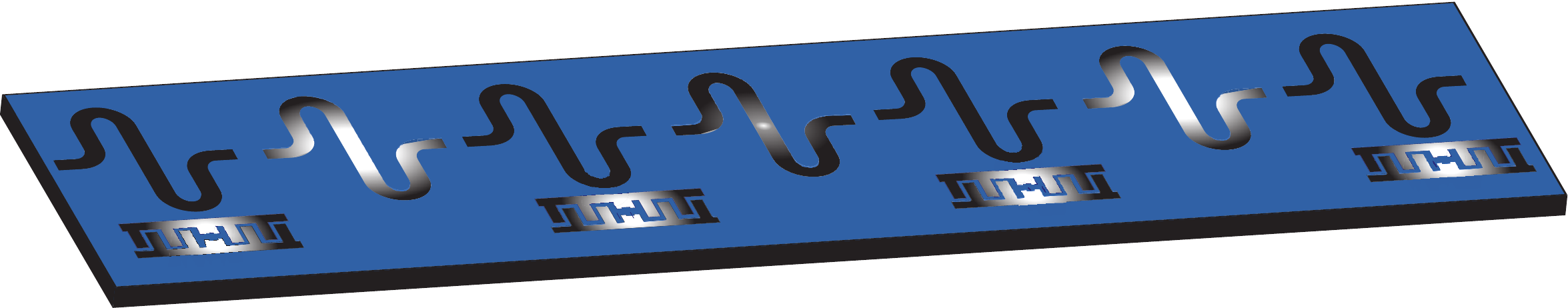}
\caption{Proposal for the realization of a strongly interacting Lieb lattice using superconducting qubits. The qubits couple to every other resonator in a chain of transmission lines. The structure is similar to the micropillar array shown in Fig.~\ref{fig:pillar}. Here, the role of the A-sites is taken by superconducting qubits, which also introduce strong photonic interactions. The shading symbolizes the crystalline nature of the photonic correlations (density wave oscillations) due to the interplay between frustration and strong interactions as shown in Fig.~\ref{fig:g2}.} 
\label{fig:array}
\end{figure}
The experimental realisation of the JCHM is a challenging task due to the requirements with respect to scalability and experimental control. 
Recently, a two-site version of the JCHM has recently been realized based on a circuit QED platform \cite{Raftery2014}, which led to the observation of 
a dissipation-induced delocalization-localization transition, where light crystallises into a self-trapped state as originally proposed in \cite{Schmidt2010*2}.\\

Circuit QED devices are artificial systems combining electronic and photonic degrees of freedom \cite{Devoret1995}. Typically, they involve microwave transmission lines and resonators with nonlinear, dissipationless elements, i.e., single Josephson junctions or superconducting quantum interference devices (SQUIDs) \cite{Wallraff2004}. Furthermore, these structures are embedded in complex photonic circuits, e.g., for initialization and readout. Such superconducting quantum-electrodynamic circuits can exhibit quantum coherence on a  macroscopic scale with sizes of qubits and resonators typically ranging from hundreds of micrometers to millimeters and coherence times extending up to the 100 $\mu$sec scale. Another appeal is the wide range of interactions that can be engineered, from weak to strong and even ultra-strong light-matter coupling \cite{Devoret2007,Niemczyk2010}.\\

The Lieb lattice in (\ref{lieb}) can be engineered in a 1D circuit QED array, where sites $A$ represent superconducting qubits and sites $B,C$ represent transmission line resonators (see Fig.~\ref{fig:array}) \cite{Biondi2015}. From the perspective of a single photon nothing changes with respect to a realization, where all three sites are represented by resonators as in (\ref{lieb}): the photon can still hop between resonators with rate $J$ or it can be exchanged between resonator $B$ and a qubit on site $A$. With qubits on site $A$, the quasi-1D Hamiltonian in (\ref{lieb}) maps exactly on a variant of the one-dimensional JCHM. Clearly, the single excitation spectra of both models are identical. However, the qubit also introduces a nonlinearity into the Hamiltonian making its spectrum highly anharmonic. By including a coherent drive, the Hamiltonian is written in a frame rotating at the drive frequency $\omega_d$ as
\begin{eqnarray}
\label{jchm}
H= \bar{H}_{\rm Lieb} + \sum_i f_i (a_i + a_i^\dagger)\,,
\end{eqnarray}
where $\bar{H}_{\rm Lieb}$ is identical to (\ref{lieb}), but where the bosonic operators $(a_i,a_i^\dagger)$ are replaced by the lowering operators for a two-level system $(\sigma^-_i,\sigma^+_i)$. The on-site energies in $\bar{H}_{\rm Lieb}$ are renormalized to $\Delta_i=\omega_i-\omega_d$. Photon losses are taken into account using a Lindblad master equation for the density matrix, i.e.,
\begin{eqnarray}
\label{densmat}
\dot{\rho}=-i[H,\rho] + (\kappa/2)\sum_i \left(\mathcal{D}[b_i]\rho + \mathcal{D}[c_i]\rho  \right)
\end{eqnarray}
with the Lindblad operator $\mathcal{D}[b]\rho=2b\rho b^\dagger - b^\dagger b \rho - \rho b^\dagger b$ and the cavity decay rate $\kappa$.\\

In general, it is hard to determine the non-equilibrium steady state (NESS) from a numerical solution of the density matrix in (\ref{densmat}) as the Hilbert space of the Hamiltonian in (\ref{jchm}) with dimensions ${\rm D} \times {\rm D}$ increases exponentially with the number of sites. In order to determine the NESS, one needs to diagonalize the Liouvillian of the system with dimensions ${\rm D}^2 \times {\rm D}^2$, which is obtained from a vectorization of the density matrix in (\ref{densmat}). Sparse diagonalization methods for the Liouvillian are thus limited to a few sites only.\\
\begin{figure}[t]
\centering
\includegraphics[width=0.4\textwidth,clip]{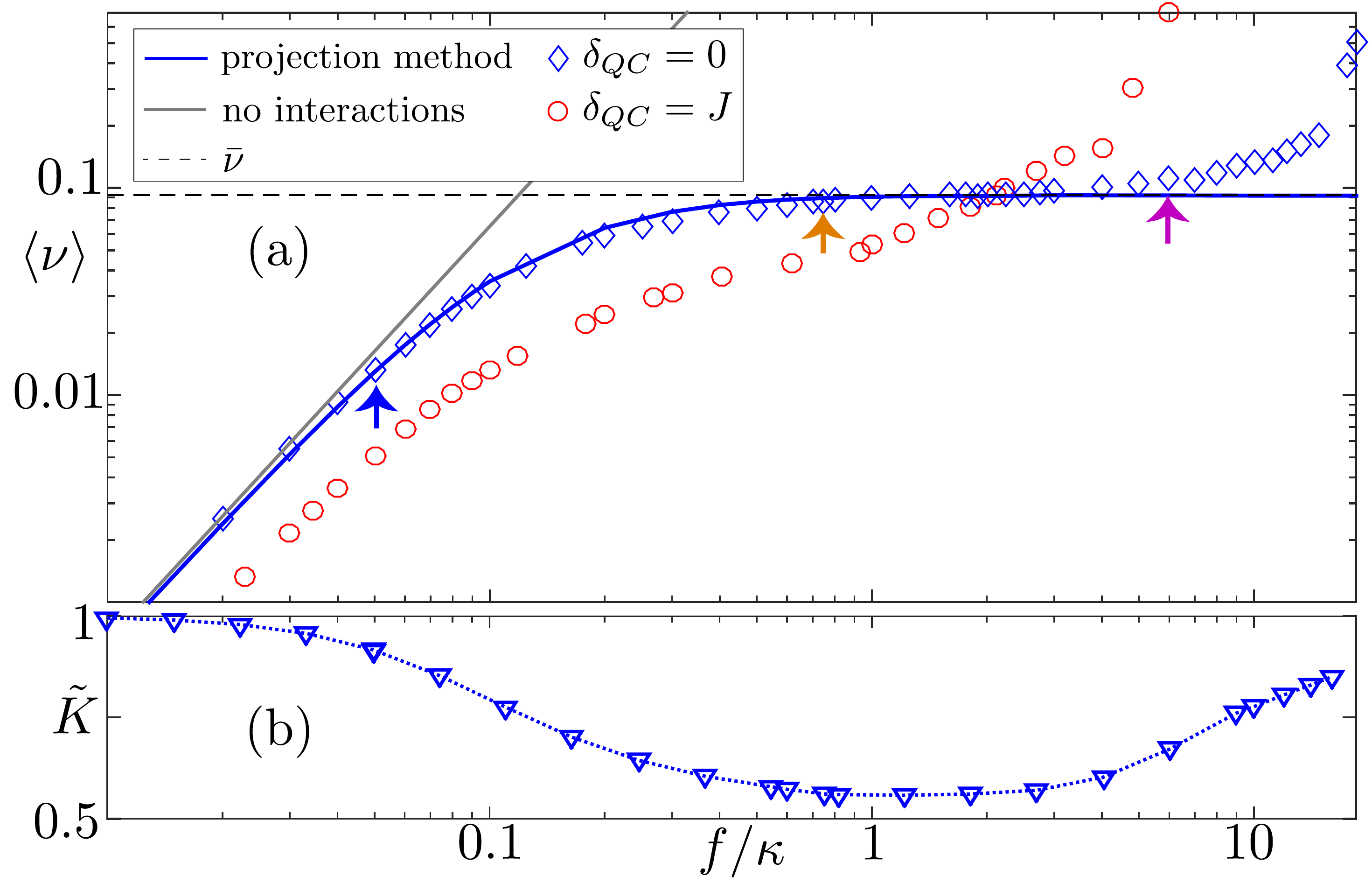}
\caption{(a) Excitation number $\langle \nu \rangle = \sum_{\rm \scriptscriptstyle X}\langle\nu_{\rm \scriptscriptstyle X} \rangle$, with $\nu_{\rm \scriptscriptstyle X} = n_{\rm \scriptscriptstyle X}/(3N)$, $n_{\rm \scriptscriptstyle X} = \sum_{j} x_j^\dagger x_j$ (X = Q, B, C and $x_j = \sigma^-_j,b_j,c_j,$) in the steady state as a function of pump strength $f/\kappa$. Shown are results obtained from projection of the density matrix on the flat band eigenspace for a system with $N=13$ unit cells and open boundary conditions (solid line) and from iTEBD simulations of the infinite system at zero detuning $\delta_{\rm \scriptscriptstyle QC}=0$ (blue symbols) and finite detuning $\delta_{\rm \scriptscriptstyle QC} = J$ (circles). The plateau at $\langle\nu\rangle\approx \bar{\nu}\approx 1/12$ is associated with a suppression of number fluctuations $\tilde{K} = [\langle (\sum_{\rm \scriptscriptstyle X} n_{\rm \scriptscriptstyle X})^2\rangle - (\sum_{\rm \scriptscriptstyle X} \langle n_{\rm \scriptscriptstyle X} \rangle)^2]/ \sum_{\rm \scriptscriptstyle X} \langle n_{\rm \scriptscriptstyle X} \rangle$ as shown in (b). Figure adapted with permission from Ref. \cite{Biondi2015}. Copyrighted by the American Physical Society.} 
\label{fig:filling}
\end{figure}
\begin{figure}[t]
\centering
\includegraphics[width=0.45\textwidth,clip]{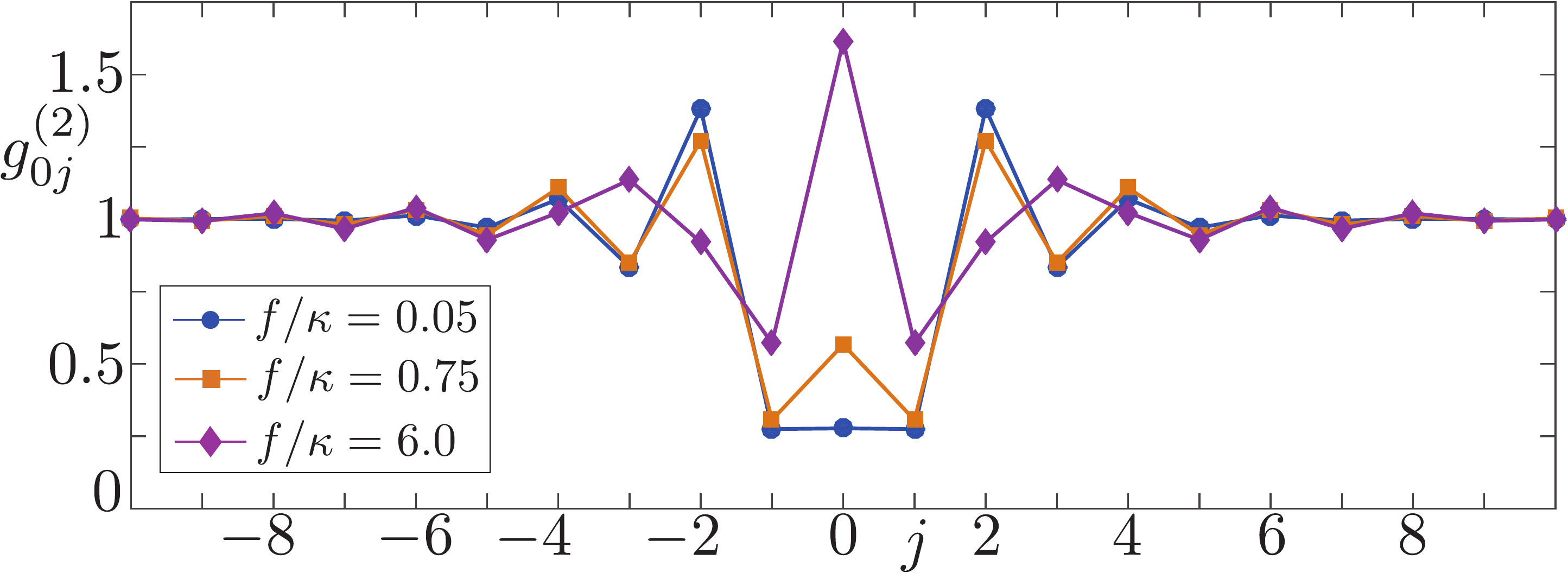}
\caption{Correlation function of photons emitted by the $C$ sites $g^{(2)}_{0j} = \langle c^\dagger_0 c^\dagger_j c_0 c_j\rangle/\langle c^\dagger_0c_0\rangle\langle c^\dagger_j c_j \rangle$ for different drive strength's $f/\kappa$ at fixed detuning $\delta_{\rm \scriptscriptstyle QC} =0$. The density-wave oscillations correspond to a period doubling with respect to the unit cell of the underlying lattice. Figure adapted with permission from Ref. \cite{Biondi2015}. Copyrighted by the American Physical Society.} 
\label{fig:g2}
\end{figure}
In the following, we discuss two alternative methods applicable in the regime of weak ($f\ll \kappa$) to moderate ($f\sim \kappa$) pump power.
For example, rather than implementing a finite-size cutoff in the number of lattice sites, one can also impose a local cutoff in the number of photons per lattice site. In this case, it is convenient to write an Ansatz for the density matrix in terms of matrix product states (MPS) \cite{Schollwoeck2011}. The time evolution of the density matrix can then be efficiently simulated for a finite as well as for the infinite system using the time-evolving block decimation algorithm (iTEBD)\cite{Vidal2007,Orus2008}.\\
The second method is based on the implementation of a many-body cutoff by projecting the density matrix only on states, which are resonant with the drive.
Those states will mostly contribute to the NESS at weak driving. Here, we also assume that the drive is resonant with the flat band, i.e., $\omega_d=\varepsilon_{\rm \scriptscriptstyle FB}$.
One can construct from (\ref{lambdastate}) exact many-particle eigenstates of $\bar{H}_{\rm Lieb}$ (for vanishing drive strength $f_i=0$) with $n>1$  by forming products of plaquettes, which do not overlap and thus have an energy that is a multiple integer of $\varepsilon_{\rm \scriptscriptstyle FB}$. For example, one can construct the two-excitation states
\begin{eqnarray}
\label{flatstates}
\ket{\Lambda_{13}} &=& \Lambda_1^\dagger\Lambda_3^\dagger\ket{\text{vac}},...,\ket{\Lambda_{1N}}=\Lambda_1^\dagger\Lambda_N^\dagger\ket{\text{vac}}\,,\nonumber\\
\ket{\Lambda_{24}} &=& \Lambda_2^\dagger\Lambda_4^\dagger\ket{\text{vac}},...\nonumber\\
\vdots\nonumber\\
{\rm etc.}\,,
\end{eqnarray}
with energy $2\varepsilon_{\rm \scriptscriptstyle FB}$. The next higher manifold is composed of the 3-excitation states $\ket{\Lambda_{135}},\ket{\Lambda_{136}},\dots$ with energy $3\varepsilon_{\rm \scriptscriptstyle FB}$. The product state with the highest filling that is still an exact eigenstate of $\bar{H}_{\rm Lieb}$ with $f_i=0$ is the density wave state
\begin{eqnarray}
\label{cdw}
\ket{\Psi_{\rm dw}} = \prod_{j=1}^{n_\text{max}}\ket{\Lambda_{2j-j_0}}
\end{eqnarray}
with energy $\varepsilon_{\rm dw} = n_\text{max}\varepsilon_{\rm \scriptscriptstyle FB}$ and filling $\nu_{\rm dw}=n_\text{max}/(3N)\approx 1/6$, where
$n_\text{max} = N/2, (N+1)/2$ denotes the particle number and $j_0 = 0,1$ is an index for $N$ even or odd respectively. Eigenstates with a higher filling belong to dispersive bands and are energetically gapped from the ladder of flat band states (otherwise they would require a double occupancy of the qubit).\\

The many body states belonging to the flat band thus form an equally spaced, bounded multi-level system indexed by the particle number $n=0,\dots,n_\text{max}$. The degeneracy of each many-body level $n$ is given by $\text{d}_n = \binom{N - n + 1}{n}$. We emphasize that it is the peculiar nature of the flat band states with zero kinetic energy (localized plaquette states), which allows us to write analytically exact many-particle eigenstates of the Hamiltonian $\bar{H}_{\rm Lieb}$ with $n>1$ and $f_i=0$. The dimension of the effective Liouvillian projected onto this flat band subspace is now considerably reduced. For example, one can conveniently calculate the steady state on a single CPU for a lattice with $13$ unit cells, i.e., $41$ sites (if the lattice terminates with one $A$ and one $Q$ site in addition to the complete 13 unit cells). In this case, the projected density matrix has dimension ${\rm D} = 610$, which allows for an exact diagonalization of the Liouvillian using sparse methods.\\

The results of the projection method and the TEBD algorithm are shown in Fig.~\ref{fig:filling}. The average filling $\langle \nu\rangle$, i.e., the photon number divided by the number of lattice sites (for a formal definition, see figure caption), displays an extended plateau as a function of drive strength, where the particle number fluctuations are strongly suppressed. This plateau can thus be interpreted as an incompressible state of photons with $\partial \langle\nu\rangle/\partial f\approx 0$. The height of the plateau does not depend on the parameters of the Hamiltonian, but
is solely determined by the geometry of the lattice. For the 1D Lieb lattice it occurs at a filling $\bar{\nu} \approx 1/12$.
This result originates in the peculiar nature of the bound, harmonic ladder of flat band states in (\ref{flatstates}). For weak pumping ($f\ll \kappa$), the system does not feel the upper bound of the harmonic ladder (given by the charge density wave state in (\ref{cdw})) and thus behaves like a harmonic oscillator (grey, solid line in Fig.~\ref{fig:filling}). At stronger pumping ($f\sim \kappa$) the system saturates as it feels the gap to the dispersive bands, which comes into play only above the charge density-wave filling ($\nu > \nu_{\rm dw}$). In this regime, all flat band states in (\ref{flatstates}) contribute almost equally to the NESS similar to a two-level system saturating half-way between ground and excited state \cite{Bishop2009}. The saturated average excitation number is thus calculated as $\bar{n} \approx (\sum_{n=0}^{n_{\text{max}}}n\,\text{d}_n)/(\sum_{n=0}^{n_{\text{max}}}\text{d}_n)= (1-1/\sqrt{5})(N/2)$ corresponding to roughly half the density-wave filling, i.e., $\bar{\nu}=\bar{n}/(3N)\approx \nu_{\rm dw}/2\approx 1/12$ (horizontal dashed line in Fig.~\ref{fig:filling}(a)).
At even stronger pumping ($f\gg \kappa$), the drive is able to overcome the gap to the dispersive bands, which then become populated as well. As a consequence the plateau is lifted and the incompressible state is destroyed. Thus, incompressibility of photons is a consequence of an unconventional photon blockade effect on a frustrated lattice, which arises from a saturation of the many-body flat band ladder \cite{Biondi2015}.\\
 
The spatial order of the photons is encoded in the second-order cross correlation function $g^{(2)}_{ij} = \langle c^\dagger_i c^\dagger_j c_i c_j\rangle/\langle c^\dagger_i c_i\rangle\langle c^\dagger_j c_j \rangle$ shown in Fig.~\ref{fig:g2}. It provides a measure for the probability to find a second photon at site $j$ if a photon at site $i$ is already present. Due to the strong coupling regime ($g\gg \kappa$), we observe local and nearest-neighbour antibunching with $g^{(2)}_{ii}, g^{(2)}_{i i+1}<1$. Interestingly, at larger distances $g^{(2)}_{ij}$ displays oscillations alternating between bunching ($g^{(2)}_{ij}>1$) and anti-bunching ($g^{(2)}_{ij}<1$). These density-wave oscillations correspond to a period-doubling of the lattice, i.e., the period of the oscillations is twice the size of one lattice unit cell (note, that we show only the $C$ sites in Fig.~\ref{fig:g2}). Thus, if the density-wave oscillations were of infinite range, they would represent a spontaneous breaking of the discrete lattice symmetry ($Z_2$ type). Such a behavior is expected in an equilibrium atomic system with the flat band as the lowest band, which is hard to engineer \cite{Aidelsburger2015}. Here, density-wave oscillations can be observed in a frustrated photonic system out of equilibrium. On the downside, only finite size oscillations are predicted (at the onset of crystallisation), because the drive mixes the pure density-wave state in (\ref{cdw}) with other flat band states at lower energies, which are also resonant with the drive frequency. Consequently, the density matrix is not in a pure state (similar to finite-temperature effects in equilibrium).\\

\section{Outlook} 

In this article, we have discussed two different examples of frustrated light-matter systems. In both cases, the interplay of geometrical frustration and the non-equilbrium nature of photons leads to non-trivial spatial structures of light. In the case of polariton condensation, the extreme sensitivity of a flat band system with respect to disorder leads to a fragmentation of the condensate \cite{Baboux2016}. In the case of a strongly nonlinear qubit-cavity array, the interplay of frustration and interactions leads to an incompressible state of photons at the onset of crystallization \cite{Biondi2015}.\\ 

Both examples were discussed in a quasi-1D Lieb lattice, which can still be simulated rather efficiently on a classical computer.
For the future, it would be fascinating to experimentally scale up these systems to two dimensions. A particular interesting subject for further investigation is the interplay of condensation and localization in a disordered array of micropillars in the presence of strong intrinsic nonlinearities. In particular, the critical properties of polariton condensates near treshhold are expected to be beyond the usually employed mean-field approximation and constitute an interesting playground for quantum simulations, e.g., for the measurement of new critical exponents in driven, dissipative systems \cite{Sieberer2013}. On the other hand, two-dimensional systems provide a natural route for the study of topological phases, e.g., by utilising artificially engineered gauge fields for photons.\\

In short, the field of frustrated photons establishes an innovative bridge between condensed matter physics and quantum optics with a wide range of potential applications ranging from material science to quantum simulations. With the first experimental realisations at hand and the exhilarating pace of technological innovations in the field of cavity/circuit QED, it holds the potential for generating major scientific insights in the near future beyond what classical computers can currently simulate.

\begin{acknowledgments}
I acknowledge fruitful discussions and collaborative work on frustrated polaritons with A. Amo, F. Baboux, M. Biondi, G. Blatter, J. Bloch, L. Ge, S. Huber, E. v. Nieuwenburg and H. E. T\"ureci. 
\end{acknowledgments}

\bibliography{bib/ref_schmidt_12_2015,bib/biblio,bib/fb_pol}

\end{document}